\documentclass[12pt,twoside]{article}
\usepackage[reqno]{amsmath}
\usepackage{amssymb}
\usepackage{times}
\usepackage{heronb}
\usepackage{epsfig}

\begin{document}

\title{Vacuum Birefringence in Strong Magnetic Fields\thanks{Presented
    at the Workshop on ``\textsc{frontier tests of quantum
      electrodynamics and physics of the vacuum}'' at Sandansky,
    Bulgaria, June 9-15, 1998}}
\authors{Walter Dittrich and Holger Gies}%
{Institut f\"ur theoretische Physik, Universit\"at T\"ubingen,\\
 Auf der Morgenstelle 14, 72076 T\"ubingen, Germany}

\maketitle

\begin{abstract}

\noindent
Table of Contents \medskip

\noindent
1. One-loop effective Lagrangian in spinor QED.

\noindent
2. Dispersion effects for low-frequency photons.

\noindent
3. Vacuum birefringence in magnetic fields.

\noindent
4. Light cone condition, effective Lagrangian approach.
\end{abstract}

\section[]{One-loop effective Lagrangian in spinor QED}

\label{1}
The purpose of the first chapter is to set the stage for light by
light scattering within the context of the effective action
approach. The original investigations can be retraced to the work of
the authors listed in Ref. \cite{1}. Here we will present a method
which is more in the spirit of Schwinger's paper \cite{2}. Essentially
all we need in order to discuss photon-photon scattering is the
expression for the vacuum persistence amplitude,

\begin{equation}
\langle 0_+|0_-\rangle =\exp \bigl\{ \text{i} W_{04} \bigr\}\,
.\label{1.1}
\end{equation}
The index 0 appended to the action indicates that no external charged
particles are in evidence, while the index 4 corresponds to the two
incoming and outgoing photons. Although the amplitude (\ref{1.1}) can
be evaluated (with some effort) for any photon energy and without
restriction to the polarization of the incoming and outgoing photons,
we want to limit ourselves to low-energy photons. The generally
non-local photon-photon interaction becomes inevitably local and the
vacuum amplitude (\ref{1.1}) can be expressed as a space-time integral
of a local Lagrange function. In particular the choice of parallel
($\|$) and perpendicular ($\bot$) polarizations of the initial photons
and of the final photons (the polarization vectors of the photons do
not change in the low-energy collision process) yield the following
amplitude:

\begin{equation}
\langle 0_+|0_-\rangle =\exp \left\{ \text{i} \int d^4x {\cal
  L}^{(1)}_{04} (x) \right\} \label{1.2}
\end{equation}
with

\begin{equation}
{\cal L}^{(1)}_{04}=: {\cal L}_\|(x)+{\cal L}_\bot (x) \label{1.3}
\end{equation}
and

\begin{equation}
{\cal L}_\|(x)=\frac{2\alpha^2}{45m^4}
({\bf E}^2-{\bf B}^2)^2\, ,\,\, {\cal L}_\bot
(x)=\frac{2\alpha^2}{45m^4} 7({\bf E\cdot B})^2\,
.\label{1.4}  
\end{equation}
Expression (\ref{1.3}-\ref{1.4}) is easily recognized as the weak
field Lagrange function as obtained from the original renormalized
Heisenberg-Euler (H.-E.) Lagrangian,

\begin{eqnarray}
{\cal L}^{(1)}_{\text{R}}({\bf E},{\bf B})
  =-\frac{1}{8\pi^2}\!
  \int\limits_0^\infty \!\frac{ds}{s^3} \text{e}^{-m^2s}
  \Biggl[ &&\!\!\!\!\!\!\!\!  (es)^2 {\cal G} \frac{\text{Re} \cosh
  \bigl(es \sqrt{2({\cal F}+\text{i} {\cal G})}\bigr)}{\text{Im} \cosh
  \bigl(es \sqrt{2({\cal F}+\text{i} {\cal G})}\bigr)}\nonumber\\
&& \left. -\frac{2}{3}(es)^2 {\cal F} -1
\right]. \label{1.5}
\end{eqnarray}
Expression (\ref{1.5}) is the result of a one-loop calculation with
external electromagnetic fields tied to all orders to the circulating
electron loop. We can make contact to our former weak-field result
(\ref{1.3}-\ref{1.4}) by expanding the integrand of (\ref{1.5})
according to 

\begin{equation}
(es)^2{\cal G} \frac{\text{Re} \cosh \%}{\text{Im} \cosh\%}
=1+\frac{2(es)^2 }{3}{\cal F} -\frac{(es)^4}{45} (4{\cal F}^2 \!+7
{\cal G}^2) +\dots \label{1.6}
\end{equation}
where

\begin{eqnarray}
{\cal F}&=&\frac{1}{4}F_{\mu\nu}F^{\mu\nu} =
  \frac{1}{2}({\bf B}^2-{\bf E}^2) \, \label{1.7}\\
{\cal G}^2&=&\left(\frac{1}{4}F_{\mu\nu}\, ^\star\! F^{\mu\nu}\right)^2 = 
  ({\bf E}\cdot{\bf B})^2\, ,\qquad ^\star\! F_{\mu\nu}= \frac{1}{2}
  \epsilon_{\mu\nu\rho\sigma}  F^{\rho\sigma}\, .\nonumber
\end{eqnarray}
After performance of the integral in (\ref{1.5}), the effective
low-energy Lagrangian (\ref{1.3}-\ref{1.4}) emerges; we rewrite it
in the form 

\begin{equation}
{\cal L}^{(1)}_{\text{R}}=\frac{2\alpha^2}{45}
\frac{1}{m^4} (4{\cal F}^2+7{\cal G}^2)\, .\label{1.8}
\end{equation}
Superimposed is the free Maxwell Lagrangian so that in the limit of
slowly varying fields, the total effective Lagrangian function reads:

\begin{eqnarray}
{\cal L}_{\text{eff}} ({\bf E},{\bf B})&=&\frac{1}{2}
  ({\bf E}^2 -{\bf B}^2) 
  +\frac{2\alpha^2}{45m^4}({\bf E}^2
  -{\bf B}^2)^2 +7\frac{2\alpha^2}{45m^4} ({\bf E}\cdot
  {\bf B})^2\, ,\label{1.9}\\
&\equiv& {\cal L}_0 +a({\bf E}^2 -{\bf B}^2)^2 +b ({\bf
  E} \cdot{\bf B})^2\, ,\label{1.10}
\end{eqnarray}
where

\begin{equation}
a\equiv \frac{2\alpha^2}{45m^4}\, ,\quad b=7a\, .\label{1.11}
\end{equation}
We shall use the information contained in (\ref{1.9}) to exhibit the
nonlinear features of quantum electrodynamics. For this reason we
compute

\begin{eqnarray}
{\bf D}&=&\frac{\partial}{\partial {\bf E}} {\cal
  L}_{\text{eff}} ({\bf E},{\bf B})\, \quad,\,\,{\bf
  H}=-\frac{\partial}{\partial {\bf B}} {\cal L}_{\text{eff}}
  ({\bf E},{\bf B})\, \label{1.12}\\ 
{\bf D}&=&{\bf \epsilon}^{\!\!\!\!\!\leftharpoonup
  \!\!\!\!\!\rightharpoonup} \cdot {\bf E}\,\qquad\qquad\,\,\,\,
  ,\,\,{\bf B}={\bf \mu}^{\!\!\!\!\!\!\leftharpoonup 
  \!\!\!\!\!\rightharpoonup} \cdot {\bf H}\, , \label{1.13}
\end{eqnarray}
which produces the electric and magnetic permeability tensors of the
vacuum \cite{1}:

\begin{eqnarray}
\epsilon_{ik}&=&\delta_{ik}+\frac{4\alpha^2}{45m^4} \bigl[
2({\bf E}^2 -{\bf B}^2) \delta_{ik} +7B_iB_k \bigr] \label{1.14}\\
\mu_{ik}&=&\delta_{ik}+\frac{4\alpha^2}{45m^4} \bigl[
2({\bf B}^2 -{\bf E}^2) \delta_{ik} +7E_iE_k \bigr]\, . \label{1.15}
\end{eqnarray}
These quantities recall the original QED-Lagrangian in which the heavy
degrees of freedom, in our case the electron fields, have been
integrated out so as to leave behind small non-linear corrections to
the linear Maxwell Lagrangian, causing interactions between
electromagnetic fields. It is precisely this phenomenon which we want
to investigate in the sequel by studying the behavior of a plane wave
field (laser light) in presence of a strong magnetic field. 

\section[]{Dispersion effects for low-frequency photons}
\setcounter{equation}{0}
Now we want to analyze dispersion effects in the low-frequency limit,
i.e., we want to study the propagation of electrodynamic waves, e.g.,
laser photons, in the presence of an external prescribed
${\bf B}$-field. For this case it is sufficient to consider the
Lagrangian of the previous chapter:

\begin{equation}
{\cal L}^{\text{eff}}=-{\cal F} +\frac{2\alpha^2}{45}
\frac{1}{m^4} \bigl[ 4{\cal F}^2 +7{\cal G}^2 \bigr]\, .\label{2.1}
\end{equation}
Since we assume our background field to be purely magnetic, we
decompose ${\bf E}$ and ${\bf B}$ according to

\begin{equation}
{\bf E}={\bf e}\, ,\qquad\,\, {\bf B}={\bf \bar{B}} +{\bf b}\,
.\label{2.2}
\end{equation}
$({\bf e},{\bf b})$ denote the plane wave field, while
${\bf \bar{B}}$ is the applied constant magnetic field. 

The constitutive equations are given by

\begin{eqnarray}
{\bf d}&=&\frac{\partial {\cal L}^{\text{eff}}}{\partial {\bf e}}\,
\qquad ,\,\,{\bf H}=-\frac{\partial {\cal L}^{\text{eff}}}{\partial
  {\bf B}} \label{2.3}\\
&=& {\bf e}+\frac{\partial {\cal L}^{(1)}}{\partial {\bf e}}\,
  ,\,\,\,\,\quad ={\bf B}-\frac{\partial {\cal L}^{(1)}}{\partial
    {\bf B}}\, .\label{2.4}
\end{eqnarray}
Linearizing these equations with respect to the wave fields ${\bf e}$
and ${\bf b}$, we find that 

\begin{equation}
{\bf d}={\bf \epsilon}^{\!\!\!\!\!\leftharpoonup
  \!\!\!\!\!\rightharpoonup} \cdot {\bf e}\,
,\qquad\,\,{\bf h}={\bf \mu}^{\!\!\!\!\!\!\leftharpoonup 
  \!\!\!\!\!\rightharpoonup\, -1} \cdot{\bf b}\, ,\label{2.5}
\end{equation}
where

\begin{eqnarray}
{\bf \epsilon}^{\!\!\!\!\!\leftharpoonup
  \!\!\!\!\!\rightharpoonup}&=&\left( 1-\frac{8\alpha^2}{45m^4}
  {\bf \bar{B}}^2 \right) 1\!\! \text{I} +\frac{28\alpha^2}{45m^4}
  {\bf \bar{B}}^2 
  ({\bf \hat{B}}\,{\bf \hat{B}})\, ,\label{2.6}\\
{\bf \mu}^{\!\!\!\!\!\!\leftharpoonup 
  \!\!\!\!\!\rightharpoonup\, -1}&=&\left( 1-\frac{8\alpha^2}{45m^4}
  {\bf \bar{B}}^2 \right) 1\!\! \text{I} -\frac{16\alpha^2}{45m^4}
  {\bf \bar{B}}^2 
  ({\bf \hat{B}}\,{\bf \hat{B}})\, ,\label{2.7}
\end{eqnarray}
with ${\bf \hat{B}}$ the applied field direction.

Since ${\bf \epsilon}^{\!\!\!\!\!\leftharpoonup
  \!\!\!\!\!\rightharpoonup}$ and
${\bf \mu}^{\!\!\!\!\!\!\leftharpoonup \!\!\!\!\!\rightharpoonup}$ are
  constants in this linear approximation, we may write the source-free
  Maxwell equations as

\begin{eqnarray}
{\bf k}\cdot{\bf d}=0\, \quad&,\quad& \quad{\bf k}\cdot
  {\bf b}=0\, 
  ,\label{2.8}\\ 
{\bf k\times e}=\omega {\bf b}\,\, &,\quad& \,\,\,{\bf k\times
  h}=-\omega {\bf d}\, ,\label{2.9} 
\end{eqnarray}
where ${\bf \hat{k}}$ denotes the direction of propagation of the
plane wave field.
Now it is convenient to distinguish between two polarization modes
where either ${\bf e}$ ($\bot$-mode) or ${\bf h}$ ($\|$-mode) points
along the direction perpendicular to the plane containing the external
magnetic field and the wave propagation direction,
(${\bf \hat{B},\hat{k}}$)-plane. Then there exist two corresponding
refractive indices $\left( \frac{|{\bf \hat{k}}|}{\omega}\right)_{
  \bot ,\|}$, which are displayed in the following together with their
respective eigenmodes:

\begin{eqnarray}
\bot -\text{mode}:&\quad&{\bf e}\equiv {\bf \hat{k}\times \hat{B}}\,
  ,\label{2.10}\\
&&{\bf h}=n_\bot\, a\,({\bf \hat{k}\cdot \hat{B}}) {\bf \hat{k}}
  -n^{-1}_\bot \, a\,{\bf \hat{B}}\, ,\nonumber\\
&&n_\bot =1+\frac{8\alpha^2}{45m^4} \bar{B}^2 \sin^2 \Theta\,
  ,\,\,\Theta=<\!\!\!) ({\bf \hat{B},\hat{k}})\nonumber\\
&&a=1-\frac{8\alpha^2}{45m^4} \bar{B}^2\, .\nonumber\\
\| -\text{mode}:&\quad&{\bf h}\equiv {\bf \hat{k}\times \hat{B}}\,
  ,\label{2.11}\\
&&{\bf e}=-n_\|\, a^{-1}\,({\bf \hat{k}\cdot \hat{B}})
  {\bf \hat{k}} +n^{-1}_\| \, a^{-1}\,{\bf \hat{B}}\, ,\nonumber\\
&&n_\| =1+\frac{14\alpha^2}{45m^4} \bar{B}^2 \sin^2 \Theta\, .
  \nonumber
\end{eqnarray}
In both eigenmodes, the plane wave field is linearly polarized. We
also see from (\ref{2.10},\ref{2.11}) that, except for propagation
along the external field direction ($\sin \Theta=0$), the vacuum
polarized by an external constant magnetic field acts like a
birefringent medium \cite{4,5}. 

All the calculations performed so far have employed the lowest-order
H.-E. Lagrangian (\ref{2.1}). However, we can improve on our previous
results concerning the permeability tensors $\epsilon_{ik},\mu_{ik}$
and the refractive indices by taking into account virtual photon
radiative corrections to the original electron loop graph. This
additional order-$\alpha$ correction yields in the weak-field limit
\cite{6}:

\begin{eqnarray}
{\cal L}^{(2)}_{\text{R}}&=&\frac{\alpha^2}{\pi m^4} \alpha \left[
  \frac{16}{81} ({\bf B}^2  -{\bf E}^2)^2 +\frac{263}{162}
  ({\bf E\cdot B})^2\right]\, ,\label{2.12}\\
&&\frac{eE}{m^2}\ll 1\, ,\quad\, \frac{eB}{m^2}\ll 1\, .\nonumber
\end{eqnarray}
This expression, together with (\ref{2.1}), leads to a modified
H.-E. effective Lagrangian:

\begin{equation}
{\cal L}_{\text{eff}}=-{\cal F} +c_1'{\cal F}^2 +c_2'{\cal G}^2
\label{2.13}
\end{equation}
\begin{equation}
\text{with}\qquad c_1'=4a'\, ,\,\quad c_2'=b' \label{2.14}
\end{equation}
\begin{equation}
\text{and}\quad\! a'=\frac{2\alpha^2}{45m^4}\left( \!1+\frac{40}{9}
  \frac{\alpha}{\pi}\! \right)\! ,\,\, b'=\frac{7\cdot
  2\alpha^2}{45m^4} 
  \left( \!1+\frac{1315}{252} \frac{\alpha}{\pi} \!\right)
  \!.\label{2.15} 
\end{equation}
The results obtained in (\ref{1.14},\ref{1.15}) now become altered
into

\begin{eqnarray}
\epsilon_{ik}&=&\delta_{ik}+2a'\,\delta_{ik}\,
2({\bf E}^2 -{\bf B}^2)  +2 b'\, B_iB_k  \nonumber\\
\mu_{ik}&=&\delta_{ik}+2a'\,\delta_{ik}\,
2({\bf B}^2 -{\bf E}^2)  +2b'\, E_iE_k \, , \label{2.16}
\end{eqnarray}
and the $n_{\bot ,\|}$ are changed according to $\left(\Theta=
  \frac{\pi}{2} \right)$:

\begin{eqnarray}
n_\bot &=&1+\frac{8\alpha^2}{45m^4}\left( 1+\frac{40}{9}
  \frac{\alpha}{\pi} \right) \bar{B}^2 \, ,\label{2.17}\\
n_\| &=&1+\frac{14\alpha^2}{45m^4}\left( 1+\frac{1315}{252}
  \frac{\alpha}{\pi} \right) \bar{B}^2 \, .\label{2.18}
\end{eqnarray}

\section[]{Vacuum birefringence in magnetic fields}
\setcounter{equation}{0}
In this chapter we want to compute the index of refraction to all
orders in the strong external field. However, we will be confining
ourselves to low-frequency photons and neglecting virtual radiative
corrections. The corresponding process can be viewed as a single
virtual electron loop with two photon vertices and an arbitrary number
of interactions with the external field. Vacuum polarization generated
by external c-number fields can best be attacked by Schwinger's
proper-time method \cite{2}. It is exactly  this strategy that
S. Adler pursues in Appendix I of his paper \cite{4}. Here it becomes
necessary to find an explicit expression for the current density
$\langle j_\mu(x)\rangle^{A}$ induced in the vacuum by the background
field $A_\mu(x)$

\begin{displaymath}
\langle j_\mu(x)\rangle^{A}=\text{i}e \,\text{tr} \bigl[ \gamma_\mu\,
G(x,x|A) \bigr]\, .
\end{displaymath}
The electron Green's function in presence of the external field $A_\mu
(x)$ is given by

\begin{displaymath}
\left[m+\gamma \left(\frac{1}{\text{i}}\partial -eA\right)\right]
G(x,x'|A) =\delta(x-x')\, .
\end{displaymath}
Thus, what is needed to calculate the induced vacuum current is the
Green's function. Since this is outlined in Adler's and Schwinger's
work, we just refer the reader to their papers -- having in mind still
another way to proceed, which will take up an appreciable fraction of
the present chapter.

It is well known that one can include vacuum polarization effects by
modifying the original Maxwell Lagrangian according to 

\begin{equation}
{\cal L}=-\frac{1}{4} F_{\mu\nu}(x)F^{\mu\nu}(x)-\frac{1}{2}
\int\!\!
d^4x'\, A^\mu(x) \,\Pi_{\mu\nu}(x,x') A^\nu(x') ,\label{3.1}
\end{equation}
where $\Pi_{\mu\nu}(x,x')$ denotes the photon polarization tensor
which describes the effect of the vacuum induced by the external
field. From (\ref{3.1}) we obtain the modified Maxwell equations; in
momentum representation  ($\bar{B}$ is now replaced by $B$),

\begin{equation}
\bigl[ k^2g_{\mu\nu} -k_\mu k_\nu+\Pi_{\mu\nu} (k,\omega,B) \bigr]
A^\nu (k) =0\, .\label{3.2}
\end{equation}
Without loss of generality we choose the ${\bf B}$-field in
$z$-direction; ${\bf k}$ is taken in the $x$-$z$ plane. The photon
polarization vector ${\bf \epsilon}$ can be resolved into

\begin{equation}
{\bf \epsilon}=a\, {\bf \epsilon}_\|+b\, {\bf \epsilon}_\bot\,
,\quad a^2+b^2=1\, ,\label{3.3}
\end{equation}
corresponding to the direction parallel and perpendicular to the plane
containing ${\bf k}$ and ${\bf B}$.

Let $\theta$ be the angle between ${\bf k}$ and ${\bf B}$; then we
have

\begin{equation}
k^\mu=\omega\, (1,\sin \Theta,0,\cos \Theta)\, ,\,\,\, \text{or}\quad
{\bf k} =\omega\, (\sin \Theta\, {\bf \hat{i}}+\cos \Theta\,
{\bf \hat{k}} ) \label{3.4}
\end{equation}
and

\begin{eqnarray}
\epsilon_\bot^\mu= (0,0,1,0)\, \qquad\qquad\,\,&,&\text{or}\quad
{\bf \epsilon}_\bot={\bf \hat{j}} \, ,\label{3.5}\\
\epsilon_\|^\mu= (0,-\cos \Theta,0,\sin \Theta)\, &,&
\text{or}\quad {\bf \epsilon}_\|=-\cos \Theta {\bf \hat{i}}+ \sin
\Theta {\bf \hat{k}}\, .\nonumber
\end{eqnarray}
If we now decompose $A^\nu (k)$ in (\ref{3.2}) according to

\begin{displaymath}
A^\nu (k)=\dots \epsilon^\nu_\bot +\dots \epsilon^\nu_\|\, ,
\end{displaymath}
and multiply equation (\ref{3.2}) from the left by ($\dots
\epsilon^\mu_\bot +\dots \epsilon^\mu_\|$), we can easily show that
the matrix elements of $\Pi_{\mu\nu}$ between $\epsilon^\mu_\|$ and
$\epsilon^\nu_\bot$ are zero, while the diagonal elements in
($\|,\bot$)-space to be considered are

\begin{equation}
\Pi_\|=\epsilon^\mu_\|\,\Pi_{\mu\nu}\,\epsilon^\nu_\|\, ,\quad
\Pi_\bot=\epsilon^\mu_\bot\,\Pi_{\mu\nu}\,\epsilon^\nu_\bot \,
.\label{3.6}
\end{equation}
The original light cone condition for the freely propagating photon,
$k^2=0$, is then modified according to 

\begin{equation}
-(k^0)^2+({\bf k}_{\| ,\bot})^2+ \Pi_{\| ,\bot}=0\, ,\label{3.7}
\end{equation}
which opens the way to discussing the vacuum as a non-linear
responding medium.

For a given frequency of the real photon, $k^0=\omega$, we obtain from
(\ref{3.7})

\begin{equation}
|{\bf k}_{\| ,\bot}| =\omega -\frac{1}{2\omega} \Pi_{\| ,\bot}\,
.\label{3.8}
\end{equation}
The complex index of refraction is defined by

\begin{equation}
\tilde{n}_{\| ,\bot}=\frac{|{\bf k}_{\| ,\bot}|}{\omega}
=1-\frac{1}{2\omega^2} \Pi_{\| ,\bot}\, .\label{3.9}
\end{equation}
We are interested in the real part of this equation which corresponds
to the index of refraction:

\begin{equation}
n_{\| ,\bot}(\omega)=1-\frac{1}{2\omega^2} \text{Re}\, \Pi_{\|
  ,\bot}\, .\label{3.10}
\end{equation}
In view of the restriction to soft laser photons, we have no interest
in the imaginary part, $\kappa=2$Im$\,(\tilde{n}\omega)$, which is
related to the absorption coefficient,

\begin{equation}
\kappa_{\| ,\bot}=-\frac{1}{\omega}\text{Im}\, \Pi_{\| ,\bot}\,
.\label{3.11}
\end{equation}
Hence we will stay in the realm below pair creation ($\omega<2m$),
where the polarization operator $\Pi_{\| ,\bot}$ must be real.

Now we turn to the evaluation of $\Pi_\|$ and $\Pi_\bot$. It is at
this stage that we make substantial use of the explicit representation
for the polarization (mass-) operator $\Pi_{\mu\nu}$ as outlined in
\cite{3} and \cite{7}. The evaluation of $\Pi_{\| ,\bot}$ can be
performed by implementing the special field configuration as indicated
in (\ref{3.4},\ref{3.5}). This then yields the following parametric
integrals:

\begin{equation}
\Pi_{\| ,\bot}=\frac{\alpha}{2\pi}\omega^2 \sin^2 \Theta
\!\int\limits_0^\infty \!\!\frac{ds}{s}\text{e}^{-\text{i}sm^2}
\!\! \!\int\limits_{-1}^{+1}\!\!\! \frac{dv}{2}\text{e}^{-\text{i}s
  \omega^2\sin^2 \Theta \!\left(\!\frac{\cos zv-\cos z}{2z\sin
      z}-\frac{1-v^2}{4}\right)} N_{\| ,\bot}\, ,  \label{3.12}
\end{equation}
where $z=esB$ and

\begin{eqnarray}
N_\|&=&-z\cot z \left( 1-v^2 +\frac{v\sin zv}{\sin z}\right)
  +z\frac{\cos zv}{\sin z} \, ,\label{3.13}\\
N_\bot&=&-\frac{z\cos zv}{\sin z} +\frac{zv\cot z \sin zv}{\sin z}
  +\frac{2z(\cos zv -\cos z)}{\sin^3 z}\, .\label{3.14}
\end{eqnarray}
Incidentally, these expressions for $N_{\| ,\bot}$ coincide with
Adler's $J_{\bot ,\|}$. 

In the sequel we need the expansions for $N_{\| ,\bot}$ for $z\ll 1$:

\begin{eqnarray}
N_\|&=&\frac{1}{2}(1-v^2)\left(1-\frac{1}{3}v^2 \right) z^2\,
  ,\label{3.15}\\
N_\bot&=&\frac{1}{2}(1-v^2)\left(\frac{1}{2}+\frac{1}{6}v^2 \right)
  z^2\, .\label{3.16}
\end{eqnarray}
Therefore, in the low-frequency $\left(\frac{\omega}{m} \sin \Theta\ll
  1\right)$ and weak-field $\left(\frac{eB}{m^2}\ll 1\right)$ limit we
obtain

\begin{eqnarray}
\Pi_{\| ,\bot}&=&\frac{\alpha}{4\pi}\omega^2\sin^2\Theta
  \int\limits_0^\infty \!\! z\, dz\, \text{e}^{-\text{i}\frac{m^2}{eB}
  z} \!\!\int\limits_0^1\! dv(1-v^2) \left(\!
  1-\frac{v^2}{3},\frac{1}{2} +\frac{v^2}{6}\! \right) \nonumber\\
&=&-\frac{\alpha}{4\pi}\omega^2 \sin^2 \Theta \left(
  \!\frac{eB}{m^2}\! \right)^2 \left[\!\left(\! \frac{28}{45}\!
  \right)_\|,
  \left(\!\frac{16}{45}\!\right)_\bot\!\right]. \label{3.17}
\end{eqnarray}
So, indeed, $\Pi_{\| ,\bot}$ turns out to be purely real and the
corresponding refractive indices are given by

\begin{equation}
n_{\| ,\bot}-1=\frac{\alpha}{4\pi}\sin^2 \Theta \left(
  \!\frac{eB}{m^2}\! \right)^2 \left[\!\left(\! \frac{14}{45}\!
  \right)_\|,
  \left(\!\frac{8}{45}\!\right)_\bot\!\right]. \label{3.18}
\end{equation}
For $\Theta=\frac{\pi}{2}$, i.e., for the photon beam orthogonal to
the ${\bf B}$-field, we have

\begin{equation}
n_{\| ,\bot}=1+\frac{\alpha}{4\pi} \left(
  \!\frac{eB}{m^2}\! \right)^2 \left[\!\left(\! \frac{14}{45}\!
  \right)_\|,
  \left(\!\frac{8}{45}\!\right)_\bot\!\right]. \label{3.19}
\end{equation}
Hence, with only one-loop calculation we obtain

\begin{equation}
\Delta n=n_\|- n_\bot =\frac{\alpha}{4\pi} \frac{6}{45} \left(
  \!\frac{eB}{m^2}\! \right)^2 \, .\label{3.20}
\end{equation}
In particular, if we consider light of wavelength $\lambda$ traversing
a path length $L$ normal to the ${\bf B}$-field, then the angular
rotation  of the plane of polarization is given by 

\begin{equation}
\Delta \varphi =\frac{1}{15} \alpha \left(
  \!\frac{eB}{m^2}\! \right)^2  \frac{L}{\lambda}\, .\label{3.21}
\end{equation}
With two-loop corrections we obtain with the aid of (\ref{2.15})

\begin{equation}
\Delta n=\frac{\alpha}{4\pi} \frac{6}{45} \left(
  \!\frac{eB}{m^2}\! \right)^2 \left( 1+\frac{25}{4}\frac{\alpha}{\pi}
  \right) \, ,\label{3.22}
\end{equation}
and (\ref{3.21}) becomes modified according to

\begin{equation}
\Delta \varphi =\frac{1}{15} \alpha \left(
  \!\frac{eB}{m^2}\! \right)^2  \frac{L}{\lambda} \left(
  1+\frac{25}{4}\frac{\alpha}{\pi} \right) \, .\label{3.23}
\end{equation}
It is the formidable task of the PVLAS experiment to detect and
measure this small angular rotation \cite{8}.

Last of all we study the limiting case of very low-energy photons in
the external constant magnetic field of arbitrary strength. Since we
are again in the region below pair creation ($\omega\ll 2m$; also
$\frac{\omega}{m} \sin \Theta\ll 1$), we can employ the polarization
operator $\Pi_{\| ,\bot}$ of (\ref{3.12}) and obtain for the index of
refraction 

\begin{equation}
n_{\| ,\bot}=1+\frac{\alpha}{4\pi} \sin^2 \Theta \, J_{\| ,\bot}\,
,\label{3.24}
\end{equation}
\begin{equation}
\text{where}\quad\! J_{\| ,\bot}=-\!\!\int\limits_0^\infty \!\!\!
\frac{dz}{z} \text{e}^{-\text{i}z \frac{m^2}{eB}} \!\! \int\limits_0^1
\!\!\! dv\, N_{\| ,\bot}(z,v)\, ,\label{3.25}
\end{equation}
with $N_{\| ,\bot}$ given in (\ref{3.13}) and (\ref{3.14}). 

The integrals can be done and lead to 

\begin{equation}
n_{\| ,\bot}=1+\frac{\alpha}{4\pi} \sin^2 \Theta \, \eta_{\| ,\bot}\,
,\label{3.26}
\end{equation}
where

\begin{eqnarray}
\eta_\|(h)&=&8\zeta'(-1,h)-4h\zeta'(0,h)-\frac{2}{3}\Psi (1+h)
  +2h\ln h-2h^2+\frac{1}{3h} -\frac{1}{3}\, ,\nonumber\\
&&\label{3.27}\\
\eta_\bot(h)&=&-4h\zeta'(0,h)+4h^2\Psi (1+h)
  -2h\ln h-2h-4h^2+\frac{2}{3}\, .\nonumber\\
&&\label{3.28}
\end{eqnarray}
$h$ stands for $\frac{m^2}{2eB}$, $\Psi$ denotes the logarithmic
derivative of the $\Gamma$-function, and $\zeta'(z,q)$ is the
derivative of the Hurwitz Zeta-function. Plots of $\eta_{\| ,\bot}$ as
well as of $n_{\| ,\bot}$ are given in the enclosed figures.

\begin{figure}
\begin{center}
\epsfig{figure=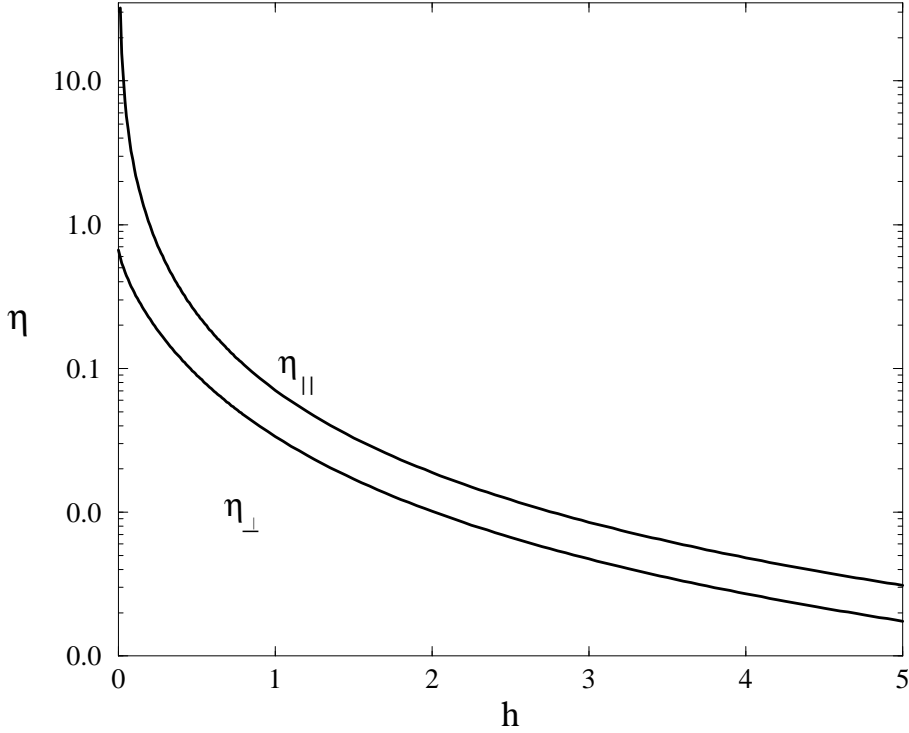,width=12cm}
\caption{$\eta_\|$ and $\eta_\bot$ in units of the dimensionless
  parameter $h=\frac{m^2}{2eB}$.}
\end{center}
\end{figure}

\begin{figure}
\begin{center}
\epsfig{figure=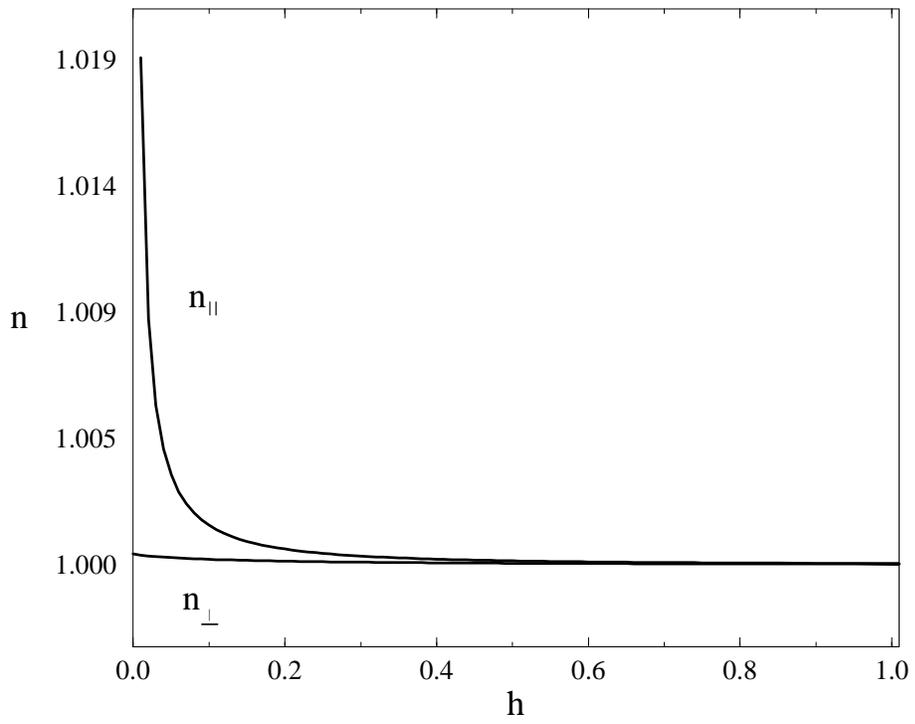,width=12cm}
\caption{Refractive indices $n_\|$ and $n_\bot$ in units of
  the dimensionless $h=\frac{m^2}{2eB}$.}
\end{center}
\end{figure}

\section[]{Light cone condition, effective Lagrangian approach}
\setcounter{equation}{0}
Let us recall that for weak electromagnetic fields we found within the
context of the H.-E. Lagrangian (\ref{2.10},\ref{2.11}):

\begin{equation}
n_\| =1+\frac{14\alpha^2}{45m^4} B^2 \sin^2 \Theta\, ,\quad
n_\bot =1+\frac{8\alpha^2}{45m^4} B^2 \sin^2 \Theta\, . 
\label{4.1}
\end{equation}
Using the relation $v_{\| ,\bot}=\frac{c(=1)}{n_{\| ,\bot}}$, we
obtain for the two phase velocities of our laser light when traversing
the external ${\bf B}$-field

\begin{equation}
v_\| =1-\frac{14\alpha^2}{45m^4} B^2 \sin^2 \Theta\, ,\quad
v_\bot =1-\frac{8\alpha^2}{45m^4} B^2 \sin^2 \Theta\, . 
\label{4.2}
\end{equation}
If we then perform an average over polarization and direction,

\begin{displaymath}
\bar{v}=\frac{1}{4\pi} \int \!d\Omega\, \frac{1}{2} (v_\|+v_\bot)\, ,
\end{displaymath}
we find

\begin{equation}
\bar{v}=1-\frac{22}{135} \frac{\alpha^2}{m^4}B^2
=1-\frac{44}{135} \frac{\alpha^2}{m^4} \, u\,.\label{4.3}
\end{equation}
Now, the authors of ref.\cite{9} claim that the velocity shift
(\ref{4.3}) holds for any (renormalized) background energy density -- 
not just electromagnetic -- with the same, i.e., universal
coefficient as in (\ref{4.3}), 

\begin{equation}
\delta\bar{v}=-\frac{44}{135} \frac{\alpha^2}{m^4} \, u\,.\label{4.4}
\end{equation}
Furthermore, G. Shore \cite{10} pointed out that the coefficients of
the velocity shifts in (\ref{4.2}) are related in a universal way to
the trace anomaly of the electromagnetic energy-momentum tensor:

\begin{equation}
\langle T^{\alpha}_{\,\,\,\alpha}\rangle_{\text{EM}}=-4 \left[\!
  \frac{8\alpha^2}{45m^4} \left(\!\frac{1}{4}
  F_{\mu\nu}F^{\mu\nu}\!\right)^2 +\frac{14\alpha^2}{45m^4}
  \bigl(\frac{1}{4}F_{\mu\nu}\, ^\star\! F^{\mu\nu}\bigr)^2 \!
  \right]. \label{4.5}
\end{equation}
It might even be possible to produce faster than light signals, e.g.,
in curved space-time manifolds or in Casimir configurations, where we
can generate velocity shifts $\delta\bar{v}>0$, since these cases
allow for negative energy densities. However, as we are solely
interested in external electromagnetic fields, we will not go into
further details concerning those other non-trivial QED vacua
\cite{11}. But already in the context of pure QED we might ask whether
the coefficients in (\ref{4.3}) or (\ref{4.4}) are really universal or
whether they are but a reflection of any weak field approach based on
the H.-E. Lagrangian. To answer this question we are now going to
introduce yet another method to set up the light cone
condition. Instead of making the electromagnetic vacuum current
\cite{4} or the polarization operator \cite{3,7} our starting point,
we want to work entirely within the effective Lagrangian
approach. Here the relevant relation turns out to be \cite{11}

\begin{equation}
k^2=Q(x,y)\, \langle T^{\mu\nu} \rangle_{x,y}\, k_\mu k_\nu\,
,\label{4.6}
\end{equation}
\begin{displaymath}
\text{with}\quad x=\frac{1}{4}F_{\mu\nu}F^{\mu\nu}\equiv {\cal F}\,
,\quad y=\frac{1}{4}F_{\mu\nu}\, ^\star\! F^{\mu\nu}\equiv {\cal G} 
\end{displaymath}
and

\begin{equation}
\langle T^{\mu\nu}\rangle_{x,y}\!=-T^{\mu\nu}_{\text{M}} \partial_x
{\cal L}(x,y) + g^{\mu\nu}\, ({\cal L}\! -x\partial_x {\cal L}\!
-y\partial_y {\cal L}).\label{4.7}
\end{equation}
${\cal L}$ is the total effective Lagrangian, i.e., Maxwell part plus
electron loop contributions. The $Q$-factor is given by

\begin{equation}
Q=\frac{\frac{1}{2} (\partial^2_x +\partial^2_y){\cal L}}
{{
\Bigl[ \!(\partial_x\!{\cal L})^2\!+(\partial_x\!{\cal
    L})\!\bigl(\!\frac{x}{2} (\partial^2_x\! -\partial^2_y)+y
  \partial_{xy}\!\bigr)\!{\cal L}\!+\frac{1}{2}\! (\partial^2_x
  \!+\partial^2_y){\cal L}(1\! -x\partial_x \!-y\partial_y)\!
  {\cal  L} \Bigr]}}. \label{4.8}
\end{equation}
In the remainder of this section we choose a special Lorentz frame,

\begin{equation}
\bar{k}^\mu=\frac{k^\mu}{|{\bf k}|} =\left(\frac{k^0}{|{\bf k}|}
  ,{\bf \hat{k}} \right) =:(v, {\bf \hat{k}})\, ,\label{4.9}
\end{equation}
where we defined the phase velocity $v=\frac{k^0}{|{\bf k}|}$. 

For (\ref{4.6}) we then obtain

\begin{equation}
v^2=1-Q\, \langle T^{\mu\nu}\rangle \bar{k}_\mu\bar{k}_\nu\,
.\label{4.10}
\end{equation}
This equation clearly demonstrates that the light cone condition
is a generalization of the ``unified formula'' of \cite{9}.

Since the $Q$-factor depends on all variables and parameters of ${\cal
  L}$, it will naturally be neither universal nor constant. For the
applications we have in mind, $Q$ will simplify substantially so
that after averaging over propagation directions, we have for $Q
  \langle T^{00}\rangle\ll 1$:

\begin{equation}
v^2=1-\frac{4}{3}\, Q\, \langle T^{00}\rangle= 1-\frac{4}{3}\, Q\, u\,
.\label{4.11}
\end{equation}
As a first example let us check the relation between velocity shifts
and scale anomaly. For weak electromagnetic fields we can take the
original H.-E. Lagrangian:

\begin{equation}
{\cal L}=-x +c_1\, x^2 +c_2\, y^2\, ,
\label{4.12}
\end{equation}
\begin{equation}
\text{with}\quad c_1=\frac{8}{45} \frac{\alpha^2}{m^4}\, ,\quad
c_2=\frac{14}{45} \frac{\alpha^2}{m^4} \, .\label{4.13}
\end{equation}
Then we get from (\ref{4.7})

\begin{eqnarray}
\langle T^{\alpha}_{\,\,\,\alpha}\rangle&=&4 ({\cal L} -x \partial_x
{\cal L} -y \partial_y {\cal L} )\label{4.14}\\
&=&-4 \Biggl(\underbrace{\frac{8}{45}\frac{\alpha^2}{m^4}}_{\sim
  \delta v_\bot}\,x^2+\underbrace{\frac{14}{45}\frac{\alpha^2}{m^4}
  }_{\sim \delta v_\|}\, y^2\Biggr)\, . \label{4.15}
\end{eqnarray}
At this stage we want to emphasize that the result (\ref{4.15}) which
expresses the energy-momentum tensor in terms of the velocity shifts
$\delta v_{\| ,\bot}$ is strictly limited to our former weak-field
approximation and hence does not hold in general \cite{11}.

However, if we take the two-loop corrected H.-E. Lagrangian \cite{6},
we obtain instead

\begin{equation}
{\cal L}=-x +c_1'\, x^2 +c_2'\, y^2\, ,
\label{4.16}
\end{equation}
with

\begin{eqnarray}
c_1'&=&\frac{8}{45} \frac{\alpha^2}{m^4} \left(
  1+\frac{40}{9}\frac{\alpha}{\pi} \right) \label{4.17}\\
c_2'&=&\frac{14}{45} \frac{\alpha^2}{m^4} \left(
  1+\frac{1315}{252}\frac{\alpha}{\pi} \right)\, , \label{4.18}
\end{eqnarray}
and (\ref{4.15}) becomes altered correspondingly. Furthermore,
restricting ourselves still to small corrections to the Maxwell
Lagrangian, we have

\begin{equation}
Q\simeq \frac{1}{2}(\partial^2_x+\partial^2_y){\cal L}=c_1'+c_2'\,
,\label{4.19}
\end{equation}
so that (\ref{4.11}) immediately yields

\begin{equation}
v=1-\frac{4\alpha^2}{135 m^4} \left(11+\frac{1955}{36}
  \frac{\alpha}{\pi} \right) \!\left[ \frac{1}{2} \bigl(
  {\bf E}^2+{\bf B}^2 \bigr) \right]\, .\label{4.20}
\end{equation}
Hence in the one-loop part of (\ref{4.20}) we can identify the factor
$\frac{44\alpha^2}{135m^4}$ as the universal constant of the ``unified
formula'' (\ref{4.4}). However, the two-loop correction in
(\ref{4.20}) destroys this universality:
$\frac{1955}{36}\frac{\alpha}{\pi}$ is as universal as 11. 

As a final remark we might add that for arbitrary magnetic field
strength, the universality becomes totally destroyed already on the
level of a one-loop calculation -- where one finds $\left(
  h=\frac{m^2}{2eB} \right)$:

\begin{eqnarray}
v^2=1- \frac{\alpha}{\pi} \frac{\sin^2 \theta}{2} \biggl[&&
  \!\!\!\!\!\!\!\!\left(
  2h^2\!-\!\frac{1}{3}\right) \Psi (1\!+\!h)-4h\ln \Gamma (h) -3h^2
  \nonumber\\
&&-h+2h\ln 2\pi +\!\frac{1}{3}\!+4 \zeta '(-1,h) +\frac{1}{6h}
  \biggr] .\nonumber\\
&&\label{4.21}
\end{eqnarray}
The total velocity shift remains rather small and amounts to

\begin{equation}
\delta v\simeq 9.58..\cdot 10^{-5}\quad \text{for}\quad
B=B_{\text{cr}}=\frac{m^2}{e} \label{4.22}
\end{equation}
with $\frac{B}{B_{\text{cr}}}< \frac{\pi}{\alpha}\simeq 430$, as is
valid within the one-loop approximation. 

Incidentally, the one-loop result for $Q$ is

\begin{equation}
Q=\frac{\alpha}{4\pi} \frac{1}{B^2} \bigl( \eta_\| +\eta_\bot \bigr)\,
,\label{4.23}
\end{equation}
where $\eta_{\| ,\bot}$ can be looked up in (\ref{3.27},\ref{3.28}). 

Upon using the relation

\begin{displaymath}
\langle T^{\mu\nu}\rangle \bar{k}_\mu\bar{k}_\nu ={\bf B}^2-
({\bf B\cdot \hat{k}})^2+{\cal O}(\alpha) =B^2 \sin^2 \theta +{\cal
  O}(\alpha)\, ,
\end{displaymath}
we find for the polarization-summed index of refraction

\begin{displaymath}
(n^2)^{-1}=1-Q\, B^2\, \sin^2\Theta\, ,
\end{displaymath}
\begin{eqnarray}
\text{or}\quad\! n&=&1+\frac{B^2}{2}\, Q\, \sin^2\Theta \nonumber\\
&=&1+\frac{1}{2} \frac{\alpha}{4\pi} \bigl( \eta_\| +\eta_\bot
  \bigr)\, \sin^2\Theta \nonumber\\
&=&\frac{1}{2}\left(1+\frac{\alpha}{4\pi}\eta_\| \sin^2 \Theta \right)
  +\frac{1}{2}\left(1+\frac{\alpha}{4\pi}\eta_\bot \sin^2 \Theta \right)
  \nonumber\\
&=&\frac{1}{2}\bigl( n_\| +n_\bot \bigr)\, , \nonumber
\end{eqnarray}
in complete agreement with our former result (\ref{3.26}).

\begin{figure}
\begin{center}
\epsfig{figure=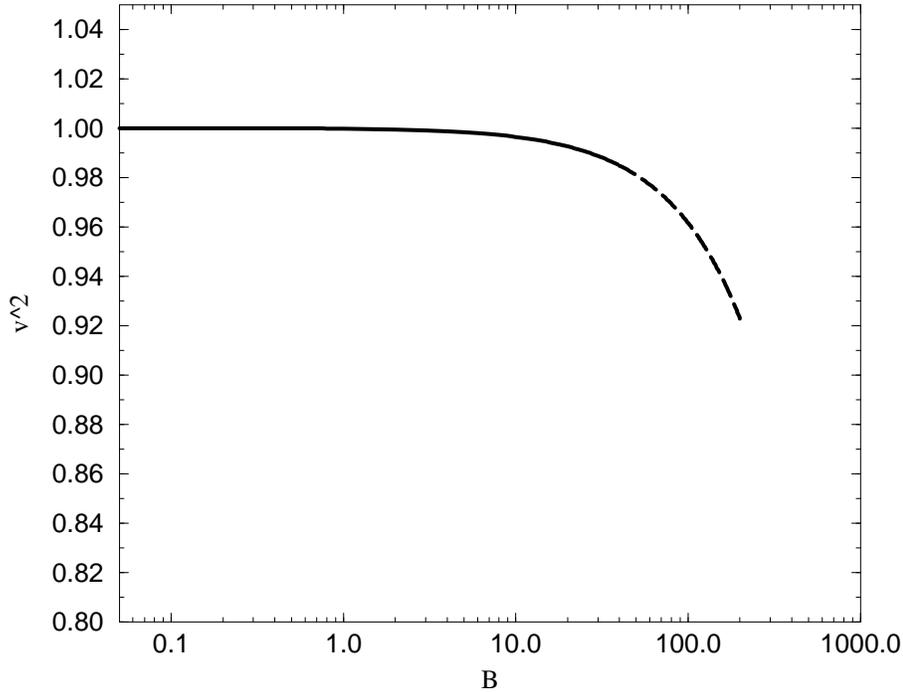,width=12cm}
\caption{Square velocity $v^2$ versus magnetic field $B$ in units of the
  critical field strength $B_{\text{cr}}=\frac{m^2}{e}$. The dashed
  curve indicates the region where higher-order corrections become
  important.} 
\end{center}
\end{figure}

\end{document}